\documentclass[aps]{revtex4}
\usepackage{graphics}
\usepackage{bm}
\newcommand{\xk}{x_\nu}
\newcommand{\xkp}{x_{\nu'}}
\newcommand{\pk}{p_\nu}
\newcommand{\pkp}{p_{\nu'}}
\newcommand{\myvec}[1]{\bm{#1}}
\newcommand{\myvect}[1]{\bm{#1}^\mathrm{T}}
\newcommand{\vj}{{\vec{j}}}
\newcommand{\vk}{{\vec{k}}}
\newcommand{\Tr}{\text{Tr}}
\newcommand{\scm}{s_\mathrm{cm}}

\newlength{\myfigwidth}
\begin{document}
%%%%%%%%%%%%%%%%%%%%%%%%%%%%%%%%%%%%%%%%%%%%%%%%%%%%%%%%%%%%%%%%%%%%%%
\title{Entanglement in the Bogoliubov vacuum} 
\author{U.~V.~Poulsen}
\affiliation{Dipartimento di Fisica, Universit\`{a} di Trento, Via
  Sommarive 14, I-38050 Povo (TN), Italy} 
\affiliation{ECT$^\star$, Strada delle Tabarelle 286, 
  I-38050 Villazzano (TN), Italy} 
\author{T.~Meyer}
\author{M.~Lewenstein} 
\affiliation{Institut f\"ur Theoretische
  Physik, Universit\"{a}t Hannover, Appelstra{\ss}e 2, 
  D-30167 Hannover, Germany}
\begin{abstract}
  We analyze the entanglement properties of the Bogoliubov vacuum,
  which is obtained as a second order approximation to the ground
  state of an interacting Bose--Einstein condensate. We work on one
  and two dimensional lattices and study the entanglement between two
  groups of lattice sites as a function of the geometry of the
  configuration and the strength of the interactions. As our measure
  of entanglement we use the logarithmic negativity, supplemented by
  an algorithmic
  check~\cite{giedke01:_entan_criter_all_bipar_gauss_states} for bound
  entanglement where appropriate. The short-range entanglement is
  found to grow approximately linearly with the group sizes and to be
  favored by strong interactions. Conversely, long range entanglement
  is favored by relatively weak interactions.  Working with periodic
  boundary conditions we find some surprising finite size effects for
  the very long range entanglement. No examples of bound entanglement
  is found.
\end{abstract}

\maketitle

%%%%%%%%%%%%%%%%%%%%%%%%%%%%%%%%%%%%%%%%%%%%%%%%%%%%%%%%%%%%%%%%%%%%%%
\section{Introduction}
\label{sec:intro}
%%%%%%%%%%%%%%%%%%%%%%%%%%%%%%%%%%%%%%%%%%%%%%%%%%%%%%%%%%%%%%%%%%%%%%

In the recent years there has been a considerable interest in studies
of entanglement in quantum distributed systems. This is a newly
developing interdisciplinary field in which quantum information theory
meets atomic and molecular physics, quantum optics, condensed matter
physics, and quantum statistical physics. There are several areas in which
the role of entanglement and quantum information in distributed
systems may be studied. The first motivation has come from the
studies of quantum macroscopic and mesoscopic phenomena in atomic
physics such as Bose-Einstein condensation (BEC)
\cite{anderson95:_obser_bose_einst_conden_dilut_atomic_vapor,davis95:_bose_einst,bradley95:_eviden_bose_einst_conden_atomic}.
The group of K. Burnett was perhaps the first to study squeezing and
entanglement of quasi-particle excitations in trapped Bose
condensates, as well as characterization of dynamical quantum states
of a zero temperature
BEC~\cite{rogel-salazar02:_squeez,rogel-salazar02:_charac,rogel-salazar03:_squeez_temp}.
In fact, in multicomponent BEC the non-linear interactions can lead to a
squeezing of the collective atomic spin, opening thus a possibility of
applications of BEC for precise frequency measurements.  This line of
research was initiated by Cirac and Zoller and their
collaborators~\cite{duan02:_quant_bose_einst,soerensen01:_many_bose}.

The new impulse to study entanglement in quantum statistical systems
has come from the papers of Amico \textit{et al.}
\cite{osterloh02:_scalin} and Nielsen \cite{osborne02:_entan} and his
collaborators, who have considered scaling properties of (short range)
entanglement close to a quantum phase transition. Following up on these
studies, various spin chain models were considered by a number of authors~\cite{vidal03:_entan_critic,vidal04:_entan_second_order,amico04:_dynam_entan,syljuasen03:_entan_sym_break,gu03:_entan_xxz,haselgrove03:_quant_states,glaser03:_entan_aniso_spin,schliemann03:_entan_su2-invar,osenda03:_tunin_entan_1d,saguia03:_entan_1d_kondo,bose02:_macros_entan_jumps,wang02:_entan_bell_heisen}.
Perhaps the most interesting result obtained so far in this context
concerns the approach of the Garching group, who has not looked only at
two parts of the distributed systems by
tracing out the rest. On the contrary, Cirac and his colleagues
considered localized entanglement of the two parts by performing optimal
local measurement on the rest of the system~\cite{verstraete04:_entan_corr}.
These authors were able to show that there exists an entanglement length
diverging at quantum critical points~\cite{verstraete04:_diver_entan_leng}.

Somewhat independently O'Connor and Wootters has introduced general
studies of spin chains and rings looking for optimal conditions for
entanglement. The optimization has been relate to certain
nearest-neighbor Hamiltonian; first in an approximate
sense~\cite{oconnor01:_entan_rings,wootters02,meyer03:_entan_ring},
but later exactly \cite{wolf03:_entan_frust}.

Apart from the spin systems, harmonic chains and rings have been
studied~\endnote{Also in this case entanglement optimizations can be
  related to looking for a ground state of a certain quadratic
  Hamiltonian~\cite{wolf03:_entan_frust}.}. As we will describe below,
powerful theoretical tools are available for these systems. An
important advantage over spin systems is that entanglement can be
studied also between two subsystems containing many sites. In
Ref.~\cite{audenaert02:_entan}, Audenaert \textit{et al.} used this fact
to consider several different subsystems on one-dimensional rings
where the coupling between neighboring oscillators is only through
position operators (i.e.\ can be interpreted as ``springs''). A
general lesson learnt was that the
ground state entanglement between two subsystems decreases rapidly
with their mutual separation. Increasing the size of the two
subsystems in general increases their entanglement, but if the
``contact'' region between the subsystems is kept fixed, a finite
limiting value is eventually reached.

The reason why harmonic chains are relatively easy to treat
theoretically is that in a grand canonical description, the ground
state as well as the thermal states belong to the category of Gaussian
states. Gaussian states are very nice for investigations from a point
of view of entanglement properties, since they are completely
characterized by their first and second order correlations. This
simplification as compared to general infinite dimensional systems has
allowed a lot of results to be derived. First of all it has been shown
that if Alice and Bob have one harmonic oscillator mode each and
share a Gaussian state, the state is entangled if and only if its
partial transpose is not
positive~\cite{simon00:_criter_cont_var,duan00:_criter_cont_var}. If
Alice has one mode and Bob many, the same conclusion holds. If,
however, both parties have more than two modes each, they may share an
entangled state which has nonetheless a positive partial transpose.
For finite dimensionsal systems the existence of such states was
demonstrated by the
Horodeckis~\cite{horodecki98:_mixed_state_entan_distil}, who have
shown that these states cannot be used for any entanglement
distillation procedures of the kind introduced by Bennett \textit{et
  al.}~\cite{bennett96:_purification}. The corresponding
distillability problem for Gaussian states was solved with identical
result: it was shown that Gaussian states are distillable if and only
if their partial transpose is not
positive~\cite{giedke01:_distil_criter_gauss}. A necessary and
sufficient entanglement criterion for Gaussian states of two parties
was found soon
after~\cite{giedke01:_entan_criter_all_bipar_gauss_states}.  All these
findings allowed for remarkable results concerning classification of
Gaussian operations~\cite{giedke02:_charac_gauss}, and in particular
the proof of the fact that Gaussian states cannot be distilled using
Gaussian operations.

As exemplified by this paper, the abovementioned results
also have big practical importance, because the Gaussian states of
photons and atoms are in many cases the ones that are easily
accessible
experimentally~\cite{braunstein03:_quant_info_cont_var,furusawa98,gottesman01:_secure_distro_squeez,grosshans02:_quant_crypt_coh_states,grosshans03:_key_dis_coh_state,silberhorn02:_key_dis_entan_beams,iblisdir03:_secure_key_dis_coh_hom}.
In particular, we will here use some of the machinery for Gaussian states
to investigate the naturally occurring entanglement in harmonic chains
of a very specific kind: The ones that appear in the studies of atoms
in optical lattices. Ultra-cold bosonic atoms in optical
lattices can undergo a super-fluid to Mott insulator transition, predicted in
Ref.~\cite{jaksch98:_cold_atoms_lattic} (see
also~\cite{oosten01:_phases_lattic}), and observed by Greiner \textit{et
 al.}~\cite{markus02:_quant_mott}.  The Mott insulator state with its
regular filling is considered as an ideal initial state for quantum
information
processing~\cite{calarco00:_quant_comp_trap,brennan99:_quant_gates},
and this has made these states the subject of intensive investigations
in the recent years. We will here instead consider the super-fluid
state of the lattice bosonic gas in the intermediate regime where
interactions are present, but not completely dominant. Then
fluctuations and excitations are described by the Bogoliubov-de Gennes
equations~\cite{pitaevskii03:_bose_einst_conden}, i.e.\ are formally
just a system of coupled harmonic oscillators. We therefore find a
similar setting as in Ref.~\cite{audenaert02:_entan} albeit with a
different form of the coupling between the oscillators. As we consider
also two-dimensional lattices, we are faced with an even wider choice
of subsystems. We have chosen to focus on subsystems that each consist
of a string of contiguous sites and to vary the size, the separation,
and (in the two-dimensional case) the relative orientation of the two
strings. Finally, we also vary the one remaining physical parameter,
namely the ratio between the energy associated with tunneling between
sites and the mean-field interaction energy.

Let us briefly summarize our results. As the separation between the two
subsystems grows, their entanglement decreases and eventually
disappears entirely. This is very similar to the results for the
``spring''-chains of Ref.~\cite{audenaert02:_entan}. Using periodic
boundary conditions a peculiar finite size effect occurs and the
entanglement does not only depend on the distance between the closest parts
of the two subsystems, even when this is very well-defined. For the
one-dimensional strings we consider, the dependence of the
entanglement on their (equal) length is essentially linear up to
moderate separations. This holds even when the relative orientation of
the strings is changed. The dependence on the interaction strength is
such that the entanglement at small separations is increased for
increased interactions while the entanglement at large separations is
decreased. For a given separation, there is thus an optimal ratio of
tunneling and interaction energy with respect to maximizing the
entanglement. Somewhat surprisingly, we find no examples of bound
entanglement.

The rest of the paper is organized in the following way. First we give some
necessary background of quantum information theory in
Sec.~\ref{sec:measures}. In Sec.~\ref{sec:non-cond} we then
calculate the relevant correlations of the system. In
Sec.~\ref{sec:results} we apply the theory to this input and we
present and analyze our results.  Finally, in
Sec.~\ref{sec:discussion} we discuss our findings and their
relations to similar work.

%%%%%%%%%%%%%%%%%%%%%%%%%%%%%%%%%%%%%%%%%%%%%%%%%%%%%%%%%%%%%%%%%%%%%%
\section{Gaussian states and entanglement}
\label{sec:measures}
%%%%%%%%%%%%%%%%%%%%%%%%%%%%%%%%%%%%%%%%%%%%%%%%%%%%%%%%%%%%%%%%%%%%%%

In general, the study of entanglement in continuous variable systems
is quite demanding. The important subset Gaussian states is however
much easier to treat and for these states the entanglement properties
are fairly well
understood~\cite{giedke02:_charac_gauss,giedke01:_entan_criter_all_bipar_gauss_states}.
The Gaussian states are well known from the field of quantum optics,
since important classes like thermal states, coherent states, and
squeezed states are all Gaussian. In the field of atom optics,
Gaussian states also appear, albeit slightly less naturally since one
tends to prefer a description in terms of Fock states, i.e.\ states
with a definite particle number, when dealing with massive particles.
This point is discussed further in Sec.~\ref{sec:discussion} below.
For the moment, however, let us just assume that we have $M$ modes
described by ``position'' and ``momentum'' operators with canonical
commutation relations
\begin{equation}
  \label{eq:commu_x_p}
  \big[ \xk,\pkp \big] = i\delta_{\nu\nu'}
  \quad,\quad
  \big[ \xk,\xkp \big] =   \big[ \pk,\pkp \big] = 0.
\end{equation}
To simplify notation, it is useful to refer to $\xk$ and
$\pk$ as $r_{2\nu}$ and $r_{2\nu+1}$, respectively. Then
Eq.~(\ref{eq:commu_x_p}) can be written as
\begin{equation}
  \label{eq:r_commu}
  \big[ r_\nu, r_{\nu'} \big] = i J_{\nu\nu'}, 
\end{equation}
where $J$ is the so-called \emph{symplectic matrix}
\begin{equation}
  \label{eq:def_J}
  J
  =
  \bigoplus_{\nu}
  \left[
  \begin{array}{cc}
     0  & 1 \\
     -1 & 0
  \end{array}
  \right]_\nu
  .
\end{equation}
Gaussian states can now be defined as states for which the Wigner
characteristic function~\cite{gardiner00:_quant_noise} is Gaussian:
\begin{equation}
  \label{eq:def_char_func}
  \chi_W(\myvec{\xi})
  =
  \left\langle
    \exp\left( 
      i \myvect{\xi} \myvec{r}
    \right)
  \right\rangle
  =
  \exp\left(i\myvect{\xi}\myvec{d} 
    +\frac{1}{4}\myvect{r}\gamma\myvec{r}\right).
\end{equation}
In Eq.~(\ref{eq:def_char_func}), $\myvec{d}=\langle \myvec{r} \rangle$
is the average displacement while
\begin{equation}
  \label{eq:1}
  \gamma_{\nu\nu'} = 2\Re \langle (r_\nu-d_\nu)(r_{\nu'}-d_{\nu'}) \rangle
\end{equation}
is the covariance matrix. Note that a Gaussian state is fully determined
by specifying $\myvec{d}$ and $\gamma$. In fact, since displacements
can always be removed by local operations, only $\gamma$ is important
for the entanglement properties of a state and in the following
subsections we summarize how information about the entanglement is
extracted from it.

%%%%%%%%%%%%%%%%%%%%%%%%%%%%%%%%%%%%%%%%%%%%%%%%%%%%%%%%%%%%%%%%%%%%%%
\subsection{Logarithmic negativity}
\label{sec:logneg}

An important problem in quantum information theory is how to
meaningfully quantify the amount of entanglement in a state. Viewing
entanglement as a resource for performing tasks forbidden by classical
physics, one seeks measures that quantify to which extend a state is
useful for a certain task or how many standard resources it would take
to create the state.

Here we apply as our measure of entanglement the logarithmic
negativity~\cite{vidal02:_comput}, $E_N$. While this measure is by no
means perfect, it does have some very nice properties for our
purposes. First of all, the value of $E_N$ provides an upper bound on
the efficiency of distillation, i.e., the extraction of maximally
entangled states from a larger number of less entangled states.
Secondly, $E_N$ is an additive quantity which means that it behaves
``naturally'' when applied to more than one copy of a state. This
facilitates comparisons between systems with different numbers of
modes. Thirdly, and quite importantly, the logarithmic negativity is
computable, i.e., there is an efficient way to calculate it for the
states we are interested in.

Formally, the logarithmic negativity is defined for any bipartite
density operator $\rho$ as
\begin{equation}
  \label{eq:def_logNeg}
  E_N(\rho) = \log_2 || \rho^\mathrm{T_A} ||_1
  ,
  %\frac{ || \rho^\mathrm{T_A} ||_1-1}{2}
\end{equation}
where $\rho^\mathrm{T_A}$ is the partially transposed (w.r.t.\ Alice)
density operator and $||\cdot||_1$ denotes the trace norm. For
Gaussian states one finds~\cite{audenaert02:_entan} 
\begin{equation}
  \label{eq:logneg_covar}
  E_N
  =
  -\sum_{\nu=1}^{2M} 
  \min \left\{
    0,\log_2\big[|\lambda_\nu(iJ\tilde{\gamma})|\big]
  \right\}
  ,
\end{equation}
where $\lambda_k(iJ\tilde{\gamma})$ denotes the $k$th eigenvalue of
$iJ\tilde{\gamma}$ and $\tilde{\gamma}$ is the covariance matrix of
$\rho^\mathrm{T_A}$. On the level of covariance matrices, partial
transposition is implemented by reversing all momenta belonging to
Alice's subsystem.

%%%%%%%%%%%%%%%%%%%%%%%%%%%%%%%%%%%%%%%%%%%%%%%%%%%%%%%%%%%%%%%%%%%%%%
\subsection{Checking for bound entanglement}
\label{sec:nonlin_alg}

One fundamental drawback of $E_N$ is the fact that a value of zero
does not guarantee that the examined state is separable: There exist
entangled density matrices with positive partial
transpose~\cite{horodecki98:_mixed_state_entan_distil,werner01:_bound_entan_gauss_states}.
To complete the picture and check whether a given state with vanishing
$E_N$ is indeed separable, we apply a qualitative test in the form of
a non-linear algorithm devised by Giedke \textit{et
al.}~\cite{giedke01:_entan_criter_all_bipar_gauss_states}. The idea of
this test is to start from the covariance matrix of the state and then
successively create a series of new matrices that all exhibit
entanglement if and only if the original state does. At each step,
simple sufficient (but not necessary) criteria for separability and
for entanglement are applied. Quite remarkably it can be shown that a
definitive answer will come out in a finite number of steps.

%%%%%%%%%%%%%%%%%%%%%%%%%%%%%%%%%%%%%%%%%%%%%%%%%%%%%%%%%%%%%%%%%%%%%%
\section{Fluctuations in the non-condensed modes}
\label{sec:non-cond}
%%%%%%%%%%%%%%%%%%%%%%%%%%%%%%%%%%%%%%%%%%%%%%%%%%%%%%%%%%%%%%%%%%%%%%

In this section we describe the calculation of the fluctuations in the
non-condensed modes for bosonic particles on a lattice. We apply the
Bogoliubov approximation which will give a good approximation to the
ground state as long as almost all atoms are condensed. The
calculations are rather straightforward, but we give them in some
detail for completeness.

%%%%%%%%%%%%%%%%%%%%%%%%%%%%%%%%%%%%%%%%%%%%%%%%%%%%%%%%%%%%%%%%%%%%%%
\subsection{The Bose-Hubbard Hamiltonian}
\label{sec:bh-hamilton}

Our calculations start with the well-known Bose-Hubbard
Hamiltonian~\cite{jaksch98:_cold_atoms_lattic,blakie04:_wannier_bose_hubbar}.
We will work with $d$ dimensional quadratic lattices and have
\begin{equation}
  \label{eq:BH-hamilton}
  H_\mathrm{BH}
  = 
  2d J \sum_\vj a_\vj^\dagger a_\vj
  - J \sum_{\langle \vj, \vj' \rangle} \left\{
  a_\vj^\dagger a_{\vj'} + \mathrm{h.c.}
  \right\}
  +
  \frac{g}{2} \sum_\vj a_\vj^\dagger a_\vj^\dagger a_\vj a_\vj
  .
\end{equation}
The first two terms in Eq.~(\ref{eq:BH-hamilton}) describes the
possibility for atoms to hop between neighboring sites. They are the
discrete equivalents of the kinetic energy with $2d$ the coordination
number (number of nearest neighbors) of the lattice. The last term
describes the collisional interaction of atoms occupying the same
site. 
The natural modes of the kinetic energy are plane waves and we
can rewrite
\begin{equation}
  \label{eq:BH-hamilton_kspace}
  H_\mathrm{BH}
  =
  J \sum_{\vk} c^\dagger_{\vk} c_\vk \; \epsilon_{\mathrm{kin}}(\vk)
  +
  \frac{g}{2N_s}\sum_{\vk_1,\vk_2,\vk_3} 
  c^\dagger_{\vk_1} c^\dagger_{\vk_2} c_{\vk_3} c_{\vk_1+\vk_2-\vk_3}
\end{equation}
with $\epsilon(\vk)=2\sum_{\sigma=x,y,\ldots} (1-\cos[2\pi
k_\sigma])$ the non-interacting energy associated with wave vector
$\vk$. $N_s$ is the total number of sites and the plane wave
annihilation operators are defined via
\begin{equation}
  \label{eq:3}
  c_\vk=\frac{1}{\sqrt{N_s}}\sum_\vj e^{-i2\pi \vk\cdot\vj} a_\vj
  .
\end{equation}

%%%%%%%%%%%%%%%%%%%%%%%%%%%%%%%%%%%%%%%%%%%%%%%%%%%%%%%%%%%%%%%%%%%%%%
\subsection{The Bogoliubov Approximation}
\label{sec:bogo-approx}

Starting from Eq.~(\ref{eq:BH-hamilton_kspace}) we now apply the
Bogoliubov approximation, i.e. we assume that one mode, the
condensate, is macroscopically populated and develop  the Hamiltonian
to second order in the fraction of non-condensed particles. Formally,
we first make the redefinitions
\begin{eqnarray}
  \label{eq:cnum}
  c_0 
  &\rightarrow& 
  e^{-i\mu t/\hbar} \big(c_0 + \sqrt{N}\big) 
  \\
  \label{eq:ckgauge}
  c_\vk 
  &\rightarrow& 
  e^{-i\mu t/\hbar} c_k\quad\mathrm{for}\quad \vk\neq\vec{0}
  \\
  \label{eq:mu}
  H_\mathrm{BH} 
  &\rightarrow& 
  H_\mathrm{BH} - \mu \sum_\vk c^\dagger_\vk c_\vk 
  ,
\end{eqnarray}
where $N$ is the number of particles and $\mu=g n$ with $n=N/N_s$ the
density of particles. This takes care of the
macroscopic population of the condensate and its resulting
time-dependence, i.e. we now work in a frame where in the fully condensed
approximation all modes are in their vacuum state and the $\vec{k}=\vec{0}$
mode defines zero energy. When only terms of
Eq.~(\ref{eq:BH-hamilton_kspace}) with at least one factor of $N$ are kept, we
get the following quadratic Hamiltonian:
\begin{equation}
  \label{eq:H_quad}
  H_\mathrm{quad}
  =
  H_0 + \sum_{\vk \neq \vec{0}} H_\vk
  ,
\end{equation}
where
\begin{eqnarray}
  \label{eq:H0}
  H_0 
  &=&
  \frac{1}{2}gn \left(
  2c^\dagger_0 c_0 + c^\dagger_0 c^\dagger_0 + c_0 c_0
  \right)
  \\
  \label{eq:Hk}
  H_\vk 
  &=&
  \left( 
  \epsilon(\vk) + gn
  \right)
  c^\dagger_\vk c_\vk
  +
  \frac{1}{2}gn \left(
  c^\dagger_\vk c^\dagger_{-\vk} + c_\vk c_{-\vk}
  \right)
  .
\end{eqnarray}

It is well-known how to diagonalize Eq.~(\ref{eq:Hk}): the normal
modes are squeezed combinations of opposite momenta. For each pair
$(\vk,-\vk)$ we get two bosonic quasi-particle modes,
\begin{eqnarray}
  \label{eq:cplus}
  c_+(\vk)
  &=&
  \cosh \eta_\vk \; c_\vk + \sinh \eta_\vk \; c^\dagger_{-\vk}
  \\
  \label{eq:cminus}
  c_-(\vk)
  &=&
  \sinh \eta_\vk \; c^\dagger_\vk + \cosh \eta_\vk \; c_{-\vk}
  ,
\end{eqnarray}
with squeezing strength
\begin{equation}
  \label{eq:def_eta}
  e^{2\eta_\vk}=\sqrt{\frac{\lambda\;\epsilon(\vk)+2}{\lambda\;\epsilon(\vk)}}
  ,
\end{equation}
where $\lambda = J/gn$ quantifies the relative importance of
interactions. Energetically, the modes are degenerate and in the limit
of vanishing interaction they correspond to the usual plane waves.

We have taken care of all modes except for the condensate itself,
the $\vk=0$ mode. If one tries to actually calculate the squeezing of
the condensate fluctuations it is found to be infinite since
$\epsilon(\vec{0})=0$. This expresses the fact that the true eigenstates of
the system must have a well-defined number of particles and therefore
the displaced vacuum assumption of Eq.~(\ref{eq:cnum}) cannot be
stationary~\cite{lewenstein96:_quant_phase_diffus_bose_einst_conden,castin98:_low_bose_einst}.
The best we can do at this point is to put in by hand a coherent state
for the condensate, i.e. to assume that the state we study is
annihilated by $c_0$ (since we have already subtracted the macroscopic
population). 

%%%%%%%%%%%%%%%%%%%%%%%%%%%%%%%%%%%%%%%%%%%%%%%%%%%%%%%%%%%%%%%%%%%%%%
\subsection{Fluctuations in a site basis}
\label{sec:site-fluc}

We now have a simple description of the system in terms of squeezed
momentum eigenstates. Since we are interested in the entanglement between
different spatial regions of the lattice, we need to calculate the
fluctuations and correlations of operators describing the atomic field
at each lattice site. To use the notation of most of quantum
information theory literature on Gaussian states, we will use
quadrature operators defined as
\begin{equation}
  \label{eq:def_x_p}
  x_\vj=\frac{ a_\vj + a^\dagger_\vj }{\sqrt{2}} 
  \quad,\quad
  p_\vj=\frac{ a_\vj - a^\dagger_\vj }{i\sqrt{2}} 
  .
\end{equation}
For the $xx$-correlations we find squeezing,
\begin{equation}
  \label{eq:xx-corr}
  \langle x_\vj x_{\vj'}\rangle-\langle x_\vj\rangle\langle x_{\vj'}\rangle
  =
  \frac{1}{2N_s}\left\{
    1 + \sum_{\vk \neq \vec{0}} 
    \cos[2\pi \vk \cdot (\vj-\vj')] e^{-2\eta_\vk}
  \right\}
  ,
\end{equation}
and for the $pp$-correlations we find anti-squeezing
\begin{equation}
  \label{eq:pp-corr}
  \langle p_\vj p_{\vj'}\rangle-\langle p_\vj\rangle\langle p_{\vj'}\rangle
  =
  \frac{1}{2N_s}\left\{
    1 + \sum_{\vk \neq \vec{0}} 
    \cos[2\pi \vk \cdot (\vj-\vj')] e^{2\eta_\vk}
  \right\}
  .
\end{equation}
In Eqs.~(\ref{eq:xx-corr}-\ref{eq:pp-corr}), the ``$1$'' is the
contribution from the condensate mode, i.e., the part we have put in by
hand. There are no $xp$-correlations.

%%%%%%%%%%%%%%%%%%%%%%%%%%%%%%%%%%%%%%%%%%%%%%%%%%%%%%%%%%%%%%%%%%%%%%
\section{Results}
\label{sec:results}
%%%%%%%%%%%%%%%%%%%%%%%%%%%%%%%%%%%%%%%%%%%%%%%%%%%%%%%%%%%%%%%%%%%%%%

In this section, we present our results for the logarithmic negativity
of bipartite states. We begin with results for 1D lattices in
Sec.~\ref{sec:res1d}, then move on to 2D lattices in
Sec.~\ref{sec:res2d}.

%%%%%%%%%%%%%%%%%%%%%%%%%%%%%%%%%%%%%%%%%%%%%%%%%%%%%%%%%%%%%%%%%%%%%%
\subsection{1D lattices}
\label{sec:res1d}

\begin{figure}[tbp]
  \centering
  \resizebox{\myfigwidth}{!}{
    \includegraphics{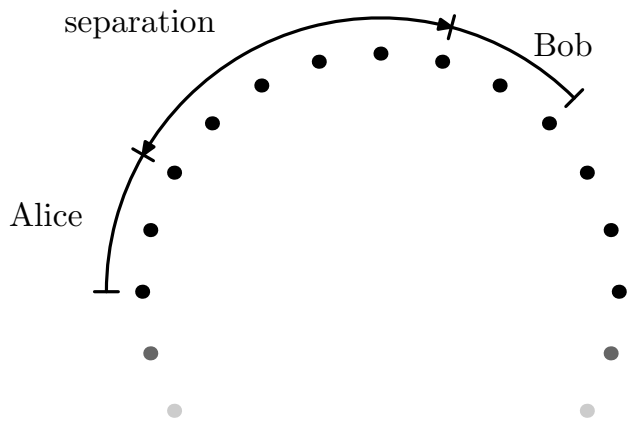}
    }  
  \caption{
    Definition of the two subsystems. Alice and Bob are both assigned
    $q$ contiguous sites and the two groups have a separation of $s$
    sites (between the extreme sites). Since we are using periodic
    boundary conditions, i.e.  working on a ring, there are in fact
    two distances between the groups: $s$ and $|N_s-2q-s|+2$. We use
    the convention that $s$ is the smaller one.}
  \label{fig:1d_setup}
\end{figure}

As we are using periodic boundary conditions, our 1D lattice is in
fact a ring. We will define the two subsystems between which we want
to study the entanglement as two sets of contiguous sites, see
Fig.~\ref{fig:1d_setup}. Because of the overall translational
invariance, only the size of the groups, $q$, and their separation,
$s$, matters.

\begin{figure}[tbp]
  \centering
  \resizebox{\myfigwidth}{!}{
    \includegraphics{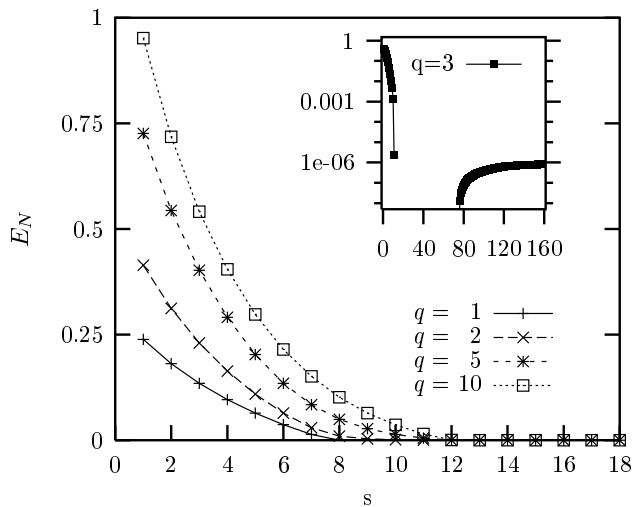}
    }
  \caption{
    Logarithmic negativity $E_N$ as a function of group separation $s$ for
    different group sizes $q$. In this plot, $N_s=321$ and $\lambda=J/gn=20$.
    When both $q$ and $s$ are much smaller than $N_s$, $E_N$ is a
    increasing function of $q$ and a decreasing function of $s$. The
    small separation part ($s < 5$) can be reasonably well approximated by an
    exponential with decay constant $\sim$ 0.3. At larger separations,
    the decay is faster. In fact, for small group sizes, $E_N$ and
    thus the distillable entanglement drops strictly to zero at quite
    moderate separations: $s=7$ for $q=1$ and $s=11$ for $q=2$. The
    insert shows a logarithmic plot for $q=3$. Here the negativity
    vanishes at $s=12$, but due to the finiteness of $N_s$, it
    reappears at $s=76$. A similar picture applies for $q=4$, while
    for larger $q$s, finite size effects always keeps $E_N$ non-zero
    for $N_s=321$.}
  \label{fig:EN_of_s_1D}
\end{figure}

In Fig.~\ref{fig:EN_of_s_1D} we plot $E_N$ as a function of the
separation of the two groups for several different group sizes. We
note that $E_N$ is generally an increasing function of the group size
and a decreasing function of the separation.
Small groups even become separable already at moderate separations: $s>6$
for $q=1$ and $s>12$ for $q=2$. Note that for $q>1$ we cannot conclude
this from the vanishing of $E_N$ alone, but we have to apply the nonlinear
algorithm described in Sec.~\ref{sec:nonlin_alg}.   

In Fig.~\ref{fig:EN_of_s_1D}, the number of sites is $N_s=321$
and the finiteness of this number naturally is important when the
separation and/or the group size is comparable to it. In that case,
there are two relevant distances from Alice to Bob: in the
clockwise direction and in the counterclockwise direction. Remarkably,
two separations that individually (i.e. as the short distance on an
infinite ring) would give rise to $E_N=0$ can still result in an
entangled state when combined. This is evidenced in the insert of
Fig.~\ref{fig:EN_of_s_1D} where the vanishing and the revival of $E_N$
as a function of $s$ for $q=3$ is plotted on a logarithmic scale. On
an infinite ring, we would have $E_N=0$ for all $s>11$, but we see that
at $s=76$, i.e.\ at a short separation of 76 and a long of
$321-2\cdot3-76=138$, $E_N$ again becomes finite. It should of course
be noted that the value $E_N$ reaches when the two groups are on
opposite sides of the ring is very low ($\sim 10^{-6}$).   

\begin{figure}[tbp]
  \centering
  \resizebox{\myfigwidth}{!}{
    \includegraphics{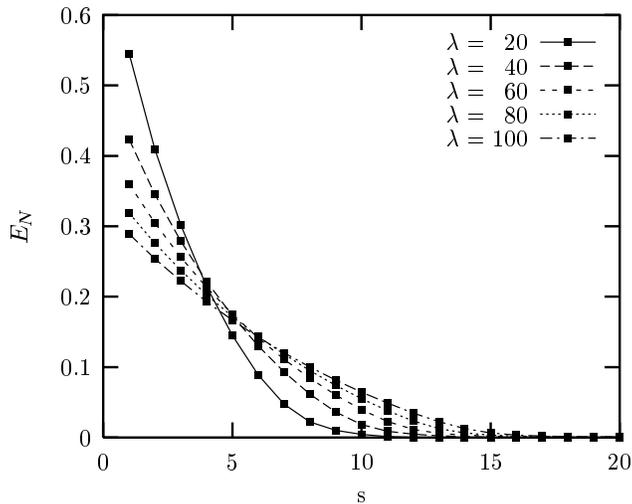}
  }
  \caption{
    Dependence of entanglement upon the value of $\lambda$. The five curves
    show $E_N$ as a function of $s$ for $q=3$ and five different
    values of $\lambda$. At short distances, a higher ratio of hopping to
    non-linearity leads to less entanglement, while at longer
    distances the opposite picture applies. }
  \label{fig:dep_lambda}
\end{figure}

Apart from the geometrical aspects regarding the lattice and the
definition of Alice's and Bob's subsystems, there is one underlying
physical parameter in the problem, namely $\lambda=J/gn$. Small values
of this parameter lead to a higher population of atoms with $\vk\neq
\vec{0}$ or, equivalently, to more squeezed quasi-particle modes. If
we keep $n$ fixed, the Bogoliubov approximation will eventually break
down and the system will enter a Mott insulator
regime~\cite{markus02:_quant_mott}.  Note, however, that since we specify only
the ratio $J/gn$, even values that lead to a high absolute number of
excited atoms are not \textit{a priori} irrelevant since the Bogoliubov
approximation still holds for a high enough $n$. In
Fig.~\ref{fig:dep_lambda} we compare the curves for $q=3$ and 
$20\le\lambda\le100$.
At short distances entanglement is clearly favored by a stronger
non-linearity (low value of $\lambda$), but perhaps a little
surprisingly, beyond separations of about 5 sites, stronger
non-linearity actually leads to less entanglement. One possible
explanation for this phenomenon can be derived from the so-called
monogamy of entanglement: Alice's and Bob's subsystems are naturally
not only entangled with each other but also with the remaining sites
in the system, in particular with the sites in the gap between them.
Since the total system is in a pure state, the mixedness of the
Alice-Bob system is exactly due to this entanglement. When the short
distance entanglement then grows due to a stronger non-linearity, the
mixedness of a highly delocalized Alice-Bob system must be expected to
increase, leaving less room for Alice-Bob
entanglement~\cite{adesso04:_deter}. To test the plausibility of this
explanation, we have examined the purity $\Tr \rho^2$
corresponding to Fig.~\ref{fig:dep_lambda}. As expected, the purity
has a stronger dependence on $\lambda$ at large separations than at
short separations, but the effect is not very pronounced and we cannot
rule out entirely different explanations of Fig.~\ref{fig:dep_lambda}.

%%%%%%%%%%%%%%%%%%%%%%%%%%%%%%%%%%%%%%%%%%%%%%%%%%%%%%%%%%%%%%%%%%%%%%
\subsection{2D lattices}
\label{sec:res2d}

\begin{figure}[tbp]
  \centering
  \resizebox{\myfigwidth}{!}{
  \includegraphics{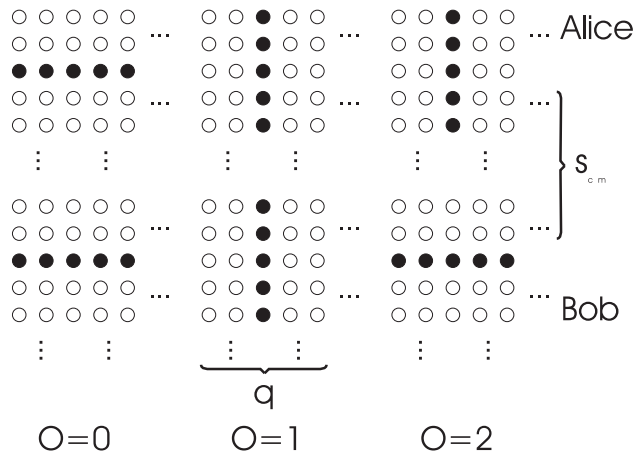}  
  }
  \caption{
    Assignments of sites to Alice and Bob on the 2D lattice. Note that
    we again use periodic boundary conditions, i.e.\ although the
    lattice is drawn as flat, we work on a torus. The ``group size''
    denoted by $q$ gives the number of sites assigned to both Alice
    and Bob. The ``center-of-mass separation'' denoted by $\scm$
    gives the distance between central sites in the two group.
    Finally, the ``orientation'' denoted by O labels different
    arrangements of the sites within each group.}
  \label{fig:2d_setup}
\end{figure}

\begin{figure}[tbp]
  \centering
  \resizebox{\myfigwidth}{!}{
  \rotatebox{-90}{
    \includegraphics{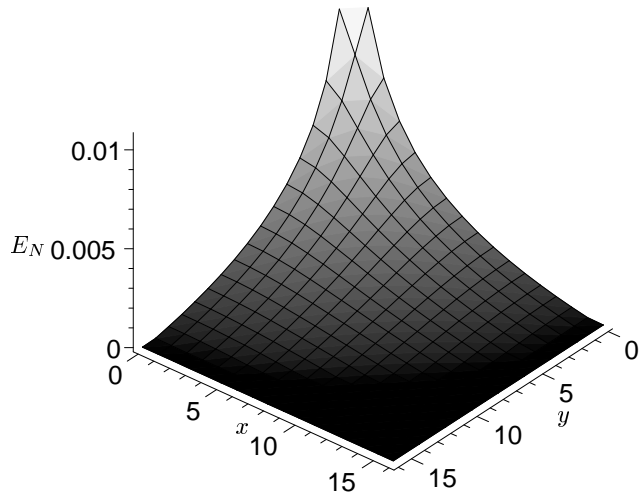}  
  }
  }
  \caption{
  Entanglement between two single sites as a function of their
  separation. The full lattice is 241 sites wide in both directions
  and $\lambda=J/gn = 100$.
  }
  \label{fig:oneone}
\end{figure}

On 2D lattices, we have a wide choice of interesting assignments of
sites to Alice and Bob. We have chosen to focus on the ones shown in
Fig.~\ref{fig:2d_setup} and will investigate to which extend the
structure within the groups influences the entanglement.
To this end, it is natural to first calculate the entanglement between
two single sites. In Fig.~\ref{fig:oneone}
we therefore show the logarithmic negativity as a function of the
separation of the sites. We note that to a good approximation $E_N$ is
an isotropic function of the separation and that it decays quickly
with increasing distance between the two sites. Compared to the 1D
results with the same $J/gn$ ratio, $E_N$ is now a factor of
approximately 10 smaller.

\begin{figure}[tbp]
  \centering
  \resizebox{\myfigwidth}{!}{
    \includegraphics{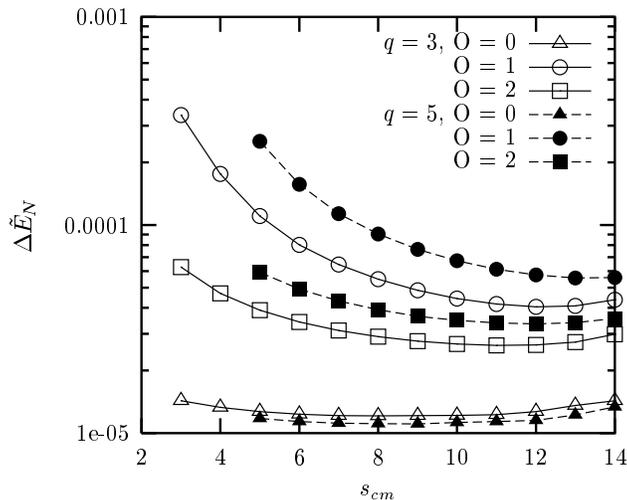}
    }
  \caption{
    Results for a 2D lattice with $\lambda=100$ and a size of 241
    sites in each dimension. Group sizes are $q=3$ (open symbols) and
    $q=5$ (filled symbols) and the three pairs of curves correspond to
    the three orientations described in Fig.~\ref{fig:2d_setup}. The
    rescaled quantity $\Delta\tilde{E}_N$ is an order of magnitude
    smaller than the single site logarithmic negativity at separations
    $\scm < 11$ indicating that the logarithmic negativity is
    approximately proportional to $q$ and that $\scm$ is a
    good measure of the distance between the groups.}
  \label{fig:2d_res}
\end{figure}

In Fig.~\ref{fig:2d_res} we plot our results for the three different
``orientations'' of the two groups. Instead of plotting directly the
logarithmic negativity, we first divide $E_N$ by the number of sites
in each group,
\begin{equation}
  \label{eq:def_Etilde}
  \tilde{E}_N (q,\scm) = \frac{1}{q}E_N(q,\scm) ,
\end{equation}
and then subtract the single site result
(Fig.~\ref{fig:oneone}) evaluated at the ``center-of-mass
distance'' (cf. Fig.~\ref{fig:2d_setup}),
\begin{equation}
  \label{eq:def_deltaEtilde}
  \Delta\tilde{E}_N (q,\scm) = \tilde{E}_N(q,\scm)-E_N(1,\scm). 
\end{equation}
This is a sensible rescaling, at least at short distances where
$\Delta \tilde{E}_N(q,\scm)$ turns out to be an order of magnitude
smaller than $E_N(1,\scm)$. It also almost collapses the curves for
group sizes $q=3$ and $q=5$ onto each other: for the orientation O=0
to a very high degree, while for O=1 and 2 some residual entanglement
remains. That the orientation O=1 where Bob's string is placed as a
continuation of Alice's string leads to the highest entanglement at
a given $\scm$ is not surprising since this is the orientation with the
shortest minimal distance between sites in the two groups.

\begin{figure}[tbp]
  \centering
  \resizebox{\myfigwidth}{!}{
    \includegraphics{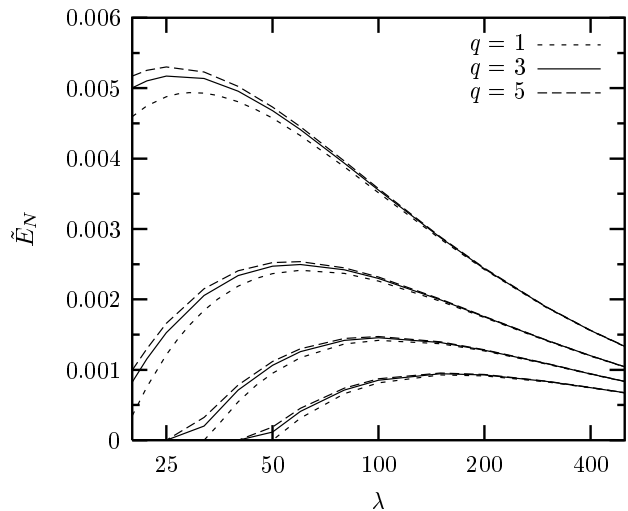}
    }
  \caption{
        Dependence of $\tilde{E}_N$ upon $\lambda$ for the orientation O=0
        on a 2D lattice (241 sites in each dimension). The dotted
        curves are single site results, $q=1$, the full curves are for  
        groups of $q=3$ sites each, and the dashed curves are for $q=5$.
        Results for 4 different separations are shown: $\scm=5,7,9$ and
        11. The three uppermost curves are for $\scm=5$, the next three for
        $\scm=7$ and so on. In the displayed $\lambda$ range, all the
        curves show a maximum, i.e. $E_N$ is not a monotone function of
        $\lambda$.}
  \label{fig:dep_lam_2d}
\end{figure}

In 1D we found a remarkable inversion in the dependence of $E_N$ on
$\lambda$ as the separation between groups were increased. A similar
effect is present in 2D as can be seen from Fig.~\ref{fig:dep_lam_2d}.
There we plot $\tilde{E}_N$ as a function of $\lambda$ for a number of
different group sizes and separations. All the curves show a maximum
in the plotted $\lambda$ range, i.e. in all cases the entanglement is
optimized at some finite ratio of hopping to non-linearity. As
expected from the successful rescaling in Fig.~\ref{fig:2d_res}, the
optimal $\lambda$ has only a very weak dependence on the group
size. Like in 1D, entanglement between groups at large distances has a
larger optimal $\lambda$ than entanglement at short distances.

%%%%%%%%%%%%%%%%%%%%%%%%%%%%%%%%%%%%%%%%%%%%%%%%%%%%%%%%%%%%%%%%%%%%%%
\section{Discussion}
\label{sec:discussion}
%%%%%%%%%%%%%%%%%%%%%%%%%%%%%%%%%%%%%%%%%%%%%%%%%%%%%%%%%%%%%%%%%%%%%%

First of all, we should comment on the symmetry breaking Bogoliubov
approach that we have applied, i.e. on the procedure of putting in
``by hand'' a coherent state for the condensate mode. It is well known
that this approach leads to identical predictions for excitation
frequencies etc.\ as the symmetry preserving approach that does not
assume superpositions of different total numbers of
particles~\cite{castin98:_low_bose_einst}. A simple connection between
the two can be made by averaging over the phase of the coherent state
leaving one with a Poissonian mixture of number states.  Instead, our
results should be seen as describing the entanglement existing between
Alice and Bob when they share knowledge of the overall phase of the
condensate. Of course, this phase cannot be regarded as a completely
classical piece of information as it can only be defined relative to
some reference
condensate~\cite{dunningham99:_phase_stand_bose_einst_conden}. For
recent investigations regarding the operational implications of
\emph{super selection rules} see e.g.\ 
Ref.~\cite{schuch04:_nonloc_super_rules} and references therein.

Turning now to our results, the success of the rescaling
(Eq.~(\ref{eq:def_Etilde})) displayed in Fig.~\ref{fig:2d_res} and
Fig.~\ref{fig:dep_lam_2d} is quite remarkable. For the configurations
we have considered, the dominant behavior of the logarithmic
negativity can be understood in terms the single site result evaluated
at $\scm$: it should simply be scaled with the number of
sites in each group. This scaling is exactly what one would expect
from the additivity of $E_N$ if the sites in the two groups paired up
in a natural way, but such a pairing is far from obvious when looking
at Fig.~\ref{fig:2d_setup}. It is an interesting question for future
studies to understand more precisely why this scaling applies. One
hint can maybe be drawn from recent studies of the asymptotic behavior
of ``spring''-chains~\cite{eisert_private}.

Another interesting conclusion we can draw is that a high degree of
squeezing (strong non-linearity) in the system does not necessarily
increase the entanglement: On the contrary, it will eventually tend to
decrease long distance entanglement. Similar results have recently
been found in related few site systems, both for the ground
state~\cite{giorda03:_groun_state_entan_inter_boson_graph} and in the
entanglement dynamics~\cite{giorda03:_mode_entan_boso_graphs}. It is
not clear whether a useful picture of the long distance behavior can
be derived from the short distance one via the principle of monogamy
of entanglement, but it would be very interesting to at least study
the bounds that exist. Since the Bogoliubov model is both relevant and
accessible, we expect our results to be useful in such studies, but
perhaps the prospects for gaining insight in this way are even greater
in quantum critical
systems~\cite{osborne02:_entan,osterloh02:_scalin}.

As a perspective, experimental measurement of the entanglement or even
implementation of quantum information processing in the system would
of course be interesting. However, in the formulation given here,
access to the full covariance matrix of the Alice-Bob system is
assumed and a reformulation in terms of more easily accessible
quantities would probably be necessary.

%%%%%%%%%%%%%%%%%%%%%%%%%%%%%%%%%%%%%%%%%%%%%%%%%%%%%%%%%%%%%%%%%%%%%%
\section*{Acknowledgements}
We gratefully acknowledge discussions with O. G\"{u}hne and support
from the EU-network \textit{Cold Quantum Gases} and the Deutsche
Forschungsgemeinschaft, GRK 282, SFB 407, SPP 1078, SPP 1116. UVP also
acknowledges support from the Danish Natural Science Research Council
and PRIN 2002 "Fault tolerance, control and stability in quantum
information precessing".
%%%%%%%%%%%%%%%%%%%%%%%%%%%%%%%%%%%%%%%%%%%%%%%%%%%%%%%%%%%%%%%%%%%%%%

% Compile with this line uncommented to produce a .bbl-file that
% should be included when submitting article:
%\bibliography{bibuvp,bogo_unpublished}

% Inserted .bbl:

%%%%%%%%%%%%%%%%%%%%%%%%%%%%%%%%%%%%%%%%%%%%%%%%%%%%%%%%%%%%%%%%%%%%%%
\end{document}